\documentclass[reprint,prb,superscriptaddress,showpacs]{revtex4-2}%

\usepackage{graphicx}
\usepackage{color}
\usepackage{amsmath}
\usepackage{amsfonts}
\usepackage{amssymb}
\usepackage{framed}
\usepackage{csquotes}
\usepackage{float}
\usepackage[normalem]{ulem}
\usepackage{xcolor}
\usepackage{soul}
\sethlcolor{green}

\providecommand{\U}[1]{\protect\rule{.1in}{.1in}}
\makeatletter
\renewcommand*{\fnum@figure}{{\normalfont\bfseries \figurename~\thefigure}}
\renewcommand*{\@caption@fignum@sep}{\textbf{ : }}
\makeatother
\usepackage{hyperref}
\hypersetup{
    colorlinks=true,
    linkcolor=blue,
    filecolor=magenta,      
    urlcolor=cyan,
}

\begin{document}

\title{Study of the Anomalous Hall effect by tuning the spin orientation in the Altermagnetic material CrSb}

\author{Sreedevi Chintalapudi}
\affiliation{Department of Physics, Birla Institute of Technology and Science Pilani Hyderabad, Telangan 500078, India}
\author{Upasana Agrawal}
\affiliation{Department of Physics, Birla Institute of Technology and Science Pilani Hyderabad, Telangan 500078, India}
\affiliation{Department of Mathematics and Computing, Birla Institute of Technology and Science Pilani Hyderabad, Telangan 500078, India}
\author{Suvadip Das}
\email{suvadip.das@hyderabad.bits-pilani.ac.in}
\affiliation{Department of Physics, Birla Institute of Technology and Science Pilani Hyderabad, Telangan 500078, India}







\begin{abstract}
Recent development in the field of altermagnetism, and increased demand for the search of applications of anomalous hall effect have ushered in a new era for novel quantum phases in materials. Quantum materials previously anticipated to be scientifically predictable have unfolded novel properties that brought them into the spotlight. These manifestations have led us to rethink our understanding of existing classification of magnetic materials and preexisting notions about anomalous hall effect in the light of topologically nontrivial phases of matter. One such recent development lies in the novel class of alter-magnetic materials with prospect for quantum computing. In this article, we delineate the spin and orbital resolved electronic spectrum, mode-decomposed phonon dispersion relations, geometrical berry curvature and topological surface states and their implications on anomalous Hall conductivity in the promising altermagnetic compound CrSb. We further utilize first principles calculations coupled with computationally efficient maximally localized wannier states of numerous magnetic configurations of the altermagnet to simulate the effect of external fields and elucidate the fact that the linear behaviour of anomalous hall conductivity with magnetization does not necessarily hold true for all magnetic classes, such as altermagnets.
\\
\end{abstract} 

\maketitle

\section{Introduction}

The quantum manifestation of Quantum Hall effect\cite{Hall, Hurd, Lorentz1892}, an effect resulting from the quantization of the electrons in a two dimensional electron gas\cite{Goer} in the presence of external magnetic fields at low temperatures, has revolutionized both worlds of quantum materials and modern day technology. With precise quantized values of hall resistances measured to integer multiples\cite{Yoshioka2002} of $h/e^{2}$ and shortly after, fractional multiples $h/fe^{2}$ leading to integer and fractional quantum hall effect in systems hosting two dimensional electron gas has ushered in novel concepts in quantum many-body physics such as composite fermions. The magnitude of hall resistance has been observed to be much larger in ferromagnets in external magnetic fields, and the additional effect of spin-orbit coupling\cite{Nagaosa2010} leading to spin accumulation at the edges results in anomalous hall effect\cite{Nagaosa2010, Sinova2015, Ong} and quantum anomalous hall effect respectively.\cite{MacDonald}\\

The technological implications of quantum anomalous Hall (QAH) \cite{WEIS200522} effect has been immense. Majorana modes \cite{Flensberg2021, Cano, Andreas}, crucial for quantum computing applications, appears to predominantly occur in topological superconductors. Chiral topological superconductors have been observed in materials with QAH-superconductor hybrid structures. Further, proximity induced superconductivity in QAH insulators have been found to induce chiral Majorana edge modes\cite{Choi2025, He2017}. Dissipation less chiral edge currents in QAH materials are promising for potential chiral spintronic devices.\\

The immense potential of Anomalous Hall effect (AHE) and Quantum anomalous Hall effect (QAHE) have led to the exploration of multiple classes of materials and their interfaces over the years with promise in electrical, optoelectronic and topological devices. This includes, but is not limited to, the double exchange mediated monotonous temperature dependence of AHE in chiral spin texture material SrRu$O_{3}$\cite{Kimbell2022}, earlier studies of AHE measurements in prototypical transition metal oxides\cite{Kondo}, spiral and helical magnetic compounds like MnSi\cite{Ishikawa, Stishov_2011}, Co-doped FeSi\cite{Morley}, Ca and Sr doped LaCoO$_{3}$\cite{Tokura, Samiolov}, heusler alloys such as Co$_{2}$CrAl\cite{Felser}, and Mn$_{5}$Ge$_{3}$\cite{Hilbert, Setti, Niu}. It has been noted that the Hall resistivity in most conventional compounds scale as $\rho_{xy}$  $\sim $ ${\rho}^{\beta}$\cite{Ong} where $\beta$ varies in between 1 and 2.\\

One of the earlier theoretical descriptions of the microscopic theory of AHE were formulated by Karplus and Luttinger (1954)\cite{Karplus}. The concept of AHE was soon connected to the geometry and topology of the electronic energy spectrum, and the geometrical berry connection was observed to correlate with the microcopic theory of AHE. This led to the detailed study and thereafter the decomposition of the AHE conductivity into three distinct contributions: intrinsic, skew-scattering and side-jump contributions\cite{Ong, MacDonald}. It was noted that the Hall resistivity $\rho_{xy}$ $\sim$ ${\rho}^2$ (intrinsic) and ${\rho}_{xy}$ $\sim$ ${\rho}$ (extrinsic) respectively.\\

With the exploration of topological materials, quantum anomalous hall effect gained momentum. QAHE is primarily observed in systems hosting non-trivial topological bands resulting in chiral edge states. QAHE, as opposed to AHE, results from broken time reversal symmetry driven by internal magnetization or local moments leading to dissipation-less edge states and associates itself with the occurence of Chern bands\cite{Jennifer}. Recent recipes for tracking altermagnets found anomalous hall effect as one of the key quantities for identifying altermagnetic compounds\cite{Dong-Hui}. Inspite of the popularlity of the field of anomalous hall conductivity, the exact dependence of the anomalous hall resistivity is still under debate.\cite{Ong}\\

Recently, a novel class of magnetism, termed \enquote{altermagnets}\cite{Libor2022, Libor2024, Igor}, has gained significant momentum. The phenomenon of altermagnetism appears in systems with opposite spin sublattices linked by crystal rotational symmetry, thereby leading to compensated magnetic order in the direct lattice space. The above presents itself with unconventional ordering of spins in the reciprocal lattice reflecting the coexistence of rotational symmetries. The spin space rotations in altermagnets can be expressed in group theoretical notation as \cite{Libor2022} $[E\parallel H] +[C_{2}\parallel AH]$, where the first term provides the symmetry transformations that interchange the atoms between two spin sublattices while the next term represents real space rotational transformations during interchange of the atoms from two sublattices. Lately, motivated by the enormous ramifications of the field, a significant effort has been invested to identify altermagnetic compounds. Some of the potential altermagnetic candidates being investigated include, but are not limited to the following materials: RuO$_{2}$\cite{Olena}, Mn$_{5}$Si$_{3}$\cite{Ishikawa}, CrSb\cite{Reimers2024}, FeSb$_{2}$\cite{Mazin}, MnF$_{2}$, CuF$_{2}$, MnO$_{2}$, MnTe\cite{Park}, LaMnO$_{3}$ and CaVO$_{3}$\cite{Gao}. The understanding of the associated symmetries of the altermagnet order parameters strengthens our understanding of the fields of magnetism and superconductivity. \\

In this paper, we provide first-principles calculations combined with maximally localized wannier functions for mapping the topological signatures in the altermagnetic compound CrSb. Further, the effect of spin orientation and the dependence of quantum anomalous hall conductivity on the magnetization for different non-collinear spin orientations have been investigated. In order to inspect the possibility of hosting surface states, calculations of slabs oriented along the (001) direction have been constructed and studied employing surface Green's functions and tight binding models. In Section II we discuss the theoretical background, basis and methodology for our first principles calculations, Section III details the findings from our calculations of anomalous hall conductivity, whereas in Section III, we delineate the conclusions and discussions following our observations in this study.

\section{Theoretical Discourse and Computational Methodology}
The Hall effect has been traditionally divided into the two main categories in metals - \textit{ordinary Hall effect} (OHE), resulting from the Lorenz force acting on electrons, and \textit{extraordinary Hall effect
}(EHE), which is a consequence of the fermionic property of electrons. Within the level of approximation for independent electrons, the OHE \cite{Hurd} is given in form of the following integral over the Fermi surface for the Hall conductivity\cite{Libor2022, smejkal2021altermagnetism, Sinova2015,Ong}:
\begin{align}
J_{x}  & =\sigma_{0}E_{x}+\sigma_{xy}(E_{y}H_{z}-E_{z}H_{y})\label{Hall}\\
\sigma_{xy}^{O}  & =-\frac{1}{12}\int\frac{dS_{F}}{|v(\mathbf{k)}|}%
\tau_{\mathbf{k}}v_{x}(\mathbf{k)}[\mathbf{v(k)\times\nabla}_{\mathbf{k}}%
]_{z}[\tau_{\mathbf{k}}v_{y}(\mathbf{k)]}\\
& =-\frac{\tau^{2}}{12}\int\frac{dS_{F}}{|v(\mathbf{k)}|}v_{x}(\mathbf{k)[}%
v_{x}\mathbf{(k)\nabla}_{y}-v_{y}\mathbf{(k)\nabla}_{x}]v_{y}(\mathbf{k)}\\
& =-\frac{\tau^{2}}{12}\int\frac{dS_{F}}{|v(\mathbf{k)}|}[v_{x}^{2}%
(\mathbf{k)}\mu_{yy}-v_{x}(\mathbf{k)}v_{y}\mathbf{(k)}\mu_{xy}],
\end{align}
where the other components are obtained by cyclic permutations. We have  utilized the isotropic relaxation time approximation in our recipe.
Here $\mu$ is the \textbf{k}-dependent invers mass tensor, defined as
\begin{equation}
\mu_{\alpha\beta}^{-1}\equiv\frac{\partial v_{\alpha}}{\partial k_{\beta}%
}\equiv\frac{\partial^{2}\epsilon_{k}}{\partial k_{\alpha}\partial k_{\beta}},
\end{equation}
and $v_{\alpha}=\partial\epsilon_{k}/dk_{\alpha}$ is the corresponding Fermi velocity and
$\epsilon_{k}$ is the electron energy respectively. If the band
energies are different for each spin direction, the summation over the spin index should be incorporated. In all our discussions we will be addressing all physical quantities in the atomic units.\\

\begin{figure*}[pt]
	\centering
	\includegraphics[width=0.7\linewidth]{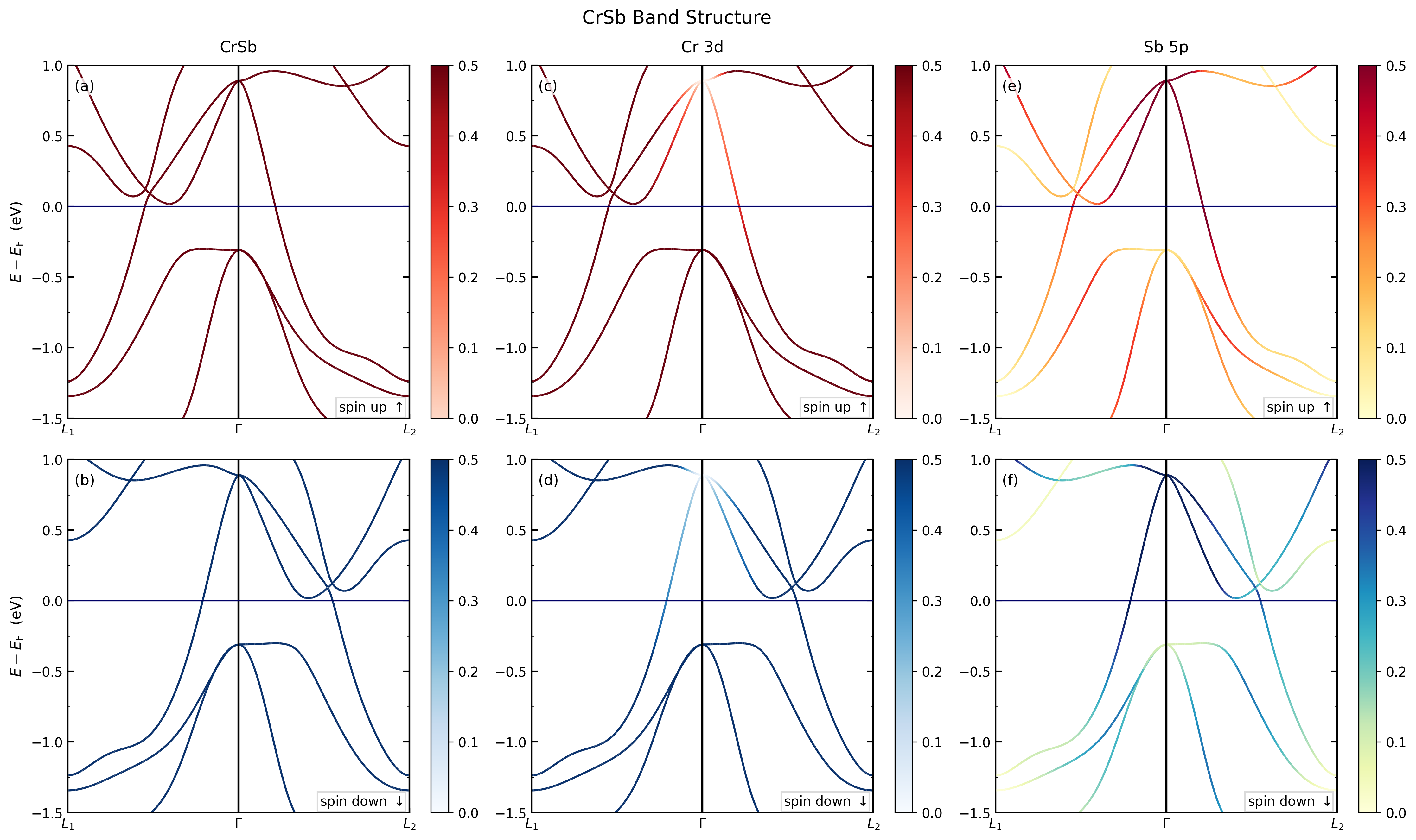}
	\caption{ Band structure of CrSb for the different spin orientations of the Cr spin moment. The results elucidate the symmetry leading to altermagnetism with separate spin polarizations of the band along particular brillouin zone directions connected by rotational symmetry.}
\end{figure*}

\begin{figure*}[pt]
    \centering
\includegraphics[width=0.5\linewidth, height=0.5\linewidth]{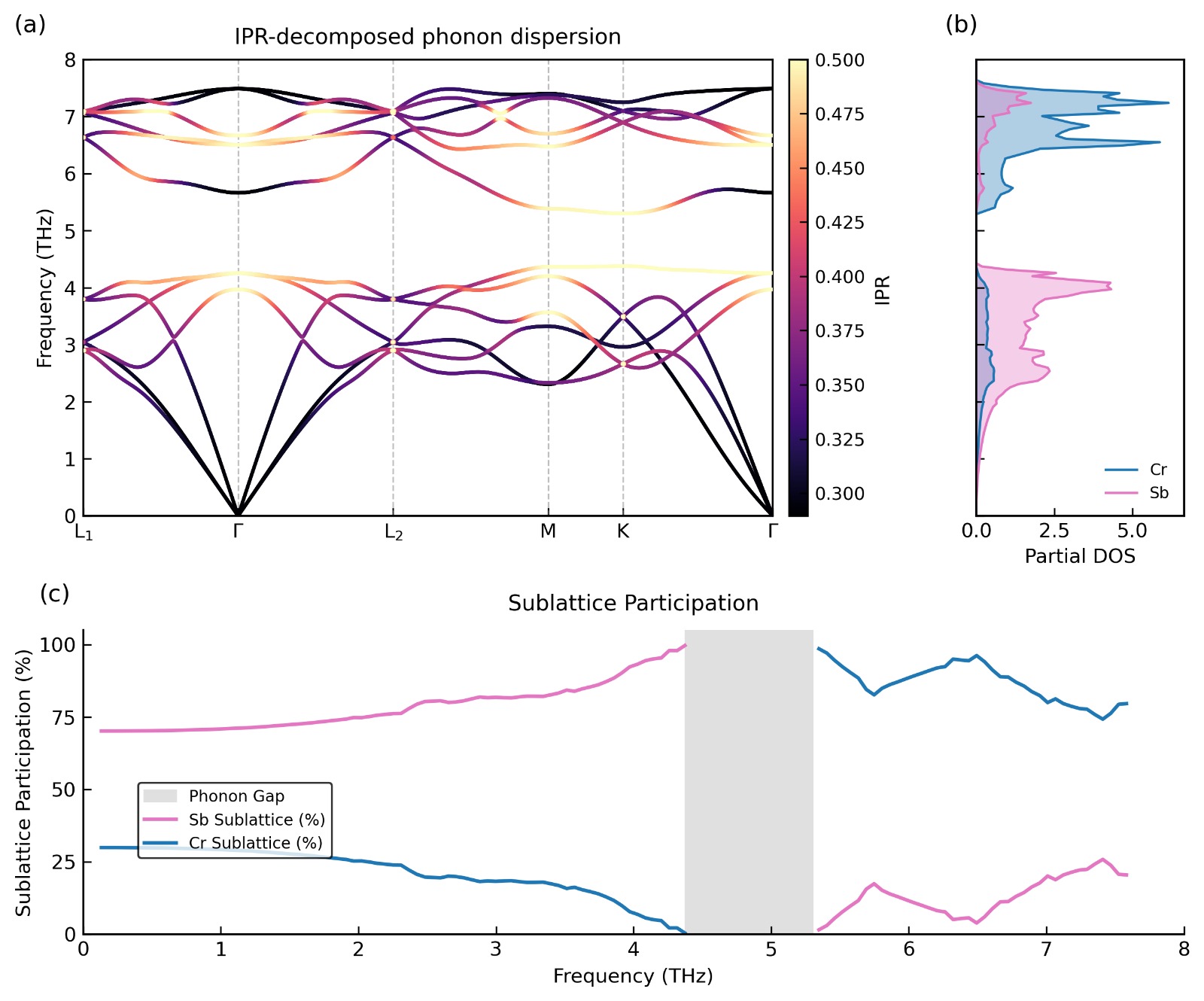}
    \caption{Phonon dispersion relations for the altermagnetic compound CrSb. (a) Phonon dispersion decomposed by Inverse participation ratio (IPR) as a function of the Brillouin zone (b) Atom decomposed phonon density of states (c) Sublattice participation fraction of phonon energy for the Cr and Sb atoms.}
    \label{fig:crsb_phonon}
\end{figure*}

From an observation of the above expression one can make following conclusions: (1) the OHE conductivity is intrinsically a property of the one-electron band structure, and, as
long as the latter does not depend on the magnetic field, OHC doesn't either (except for a small correction proportional to $H^{2}$ for a nonmagnetic material and proportional to $H$ for a ferromagnetic one, but in either case will be small and scales as the product $HN(0)$, where $N$ is the density of states) (2) as long as the OHC is small compared to the diagonal component of conductivity $\sigma_{0},$ one can invert this relation obtaining the Hall resistivity as%
\begin{equation}
R_{H}^{O}=a^{O}H,
\end{equation}
where $a$ is the coefficient determined by the one-electron nonrelativistic band structure.\\

It was discovered during the 1930s\cite{Pugh} that in the case of ferromagnets, there is another contribution to the Hall effect, which is independent of the applied magnetic field in single-domain or saturated samples, and proportional to the net
magnetization within the hysteresis term, $i.e.,$ it is a constant for a single domain, whose sign depend on the sign of the magnetization (so that both the total magnetization and the Hall conductivity are independently proportional to the imbalance of the domain population).\\

This leads to two corollaries from the above observation: (1) the effect has to be relativistic in nature, otherwise it cannot depend on the sign of magnetization (or on it direction), since spin-orbit coupling interaction decouples the spin and the spatial directions and (2) apart from the magnetic field dependence due to  domain wall dyanmics ($i.e.,$ extrinsic) there is no dependence on the applied external field. The latter appears natural since magnetic
field is small compared to the typical electron energy scale and cannot affect the electron dynamics in a saturated ferromagnet.\\

\begin{figure*}[pt]
	\centering
	\includegraphics[width=0.8\linewidth]{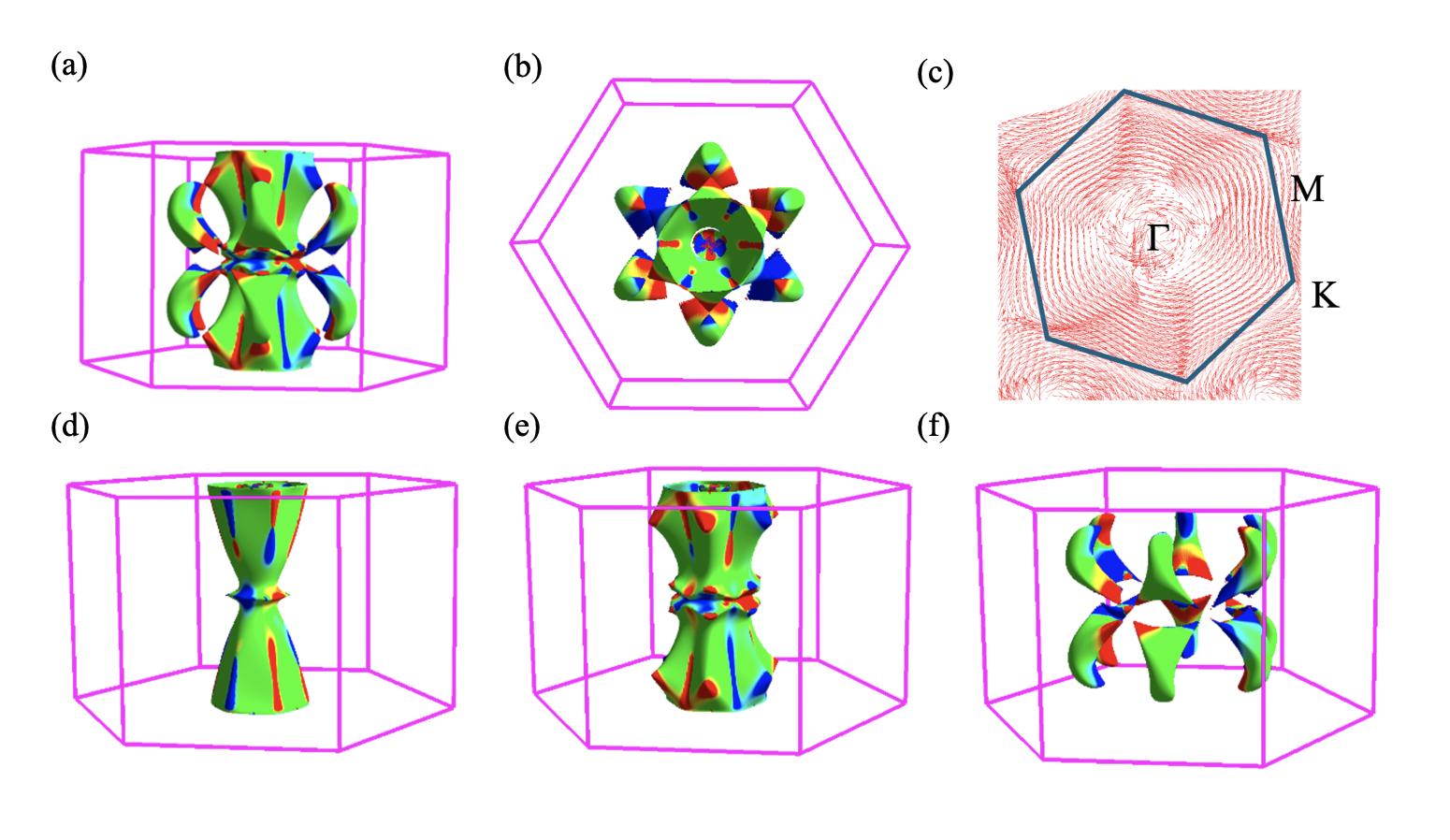}
	\caption{Fermi surface of CrSb color coded by the momentum dependent berry curvature contributions from (a) lateral view (b) top view (c) quiver plot of berry curvature in the hexagonal ab plane within the Brillouin zone (d)-(f) band decomposed berry curvature contributions to the anomalous hall conductivity.}
\end{figure*}

\begin{figure*}[pt]
	\centering
	\includegraphics[width=0.8\linewidth]{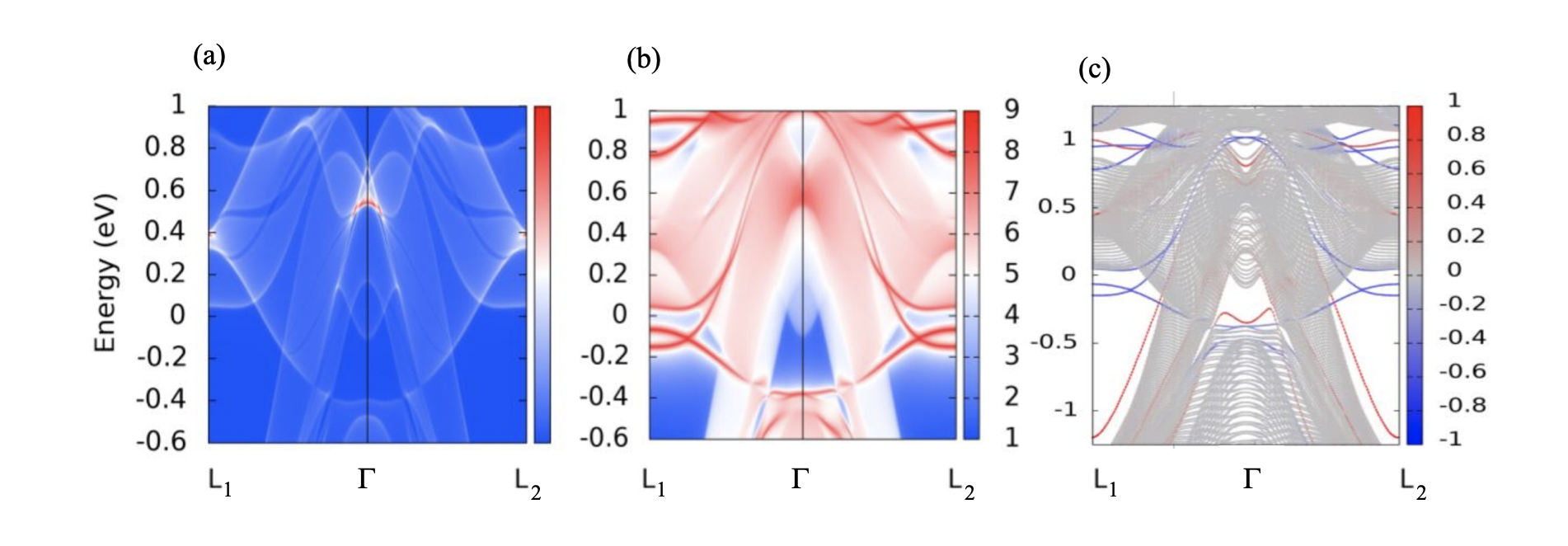}
	\caption{Slab calculations of the projected energy spectrum along the (001) plane obtained from tight binding calculations showing (a) the bulk energy spectrum (c) surface density of states and energy spectrum (c) existence of topological surface states.}
\end{figure*}

\begin{figure*}[pt]
	\centering	\includegraphics[width=1.0\columnwidth]{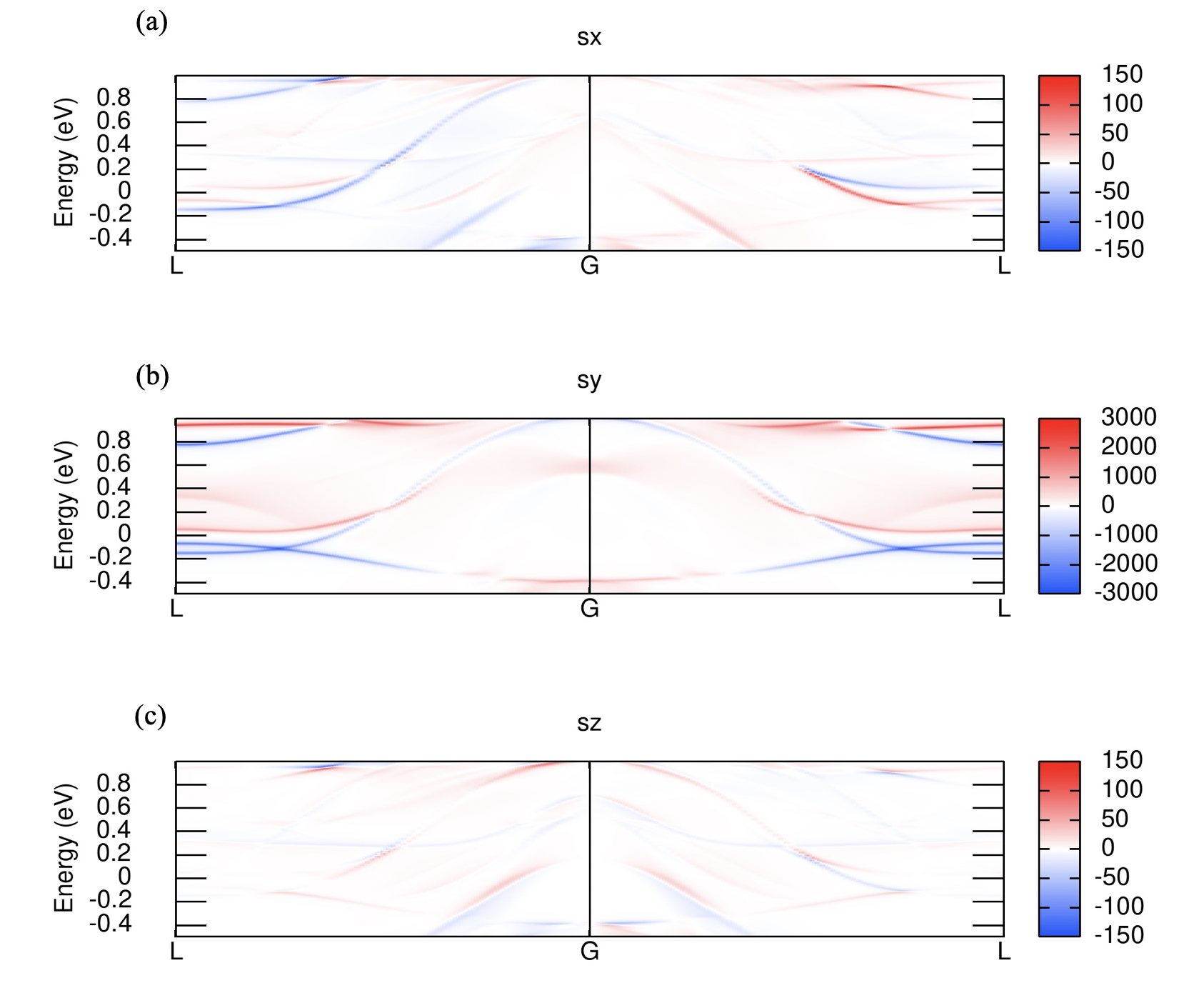}
	\caption{ The spin resolved density of states projected spectrum from the (001) slab calculations of the tight binding model showing the projected Pauli spin matrices (a) $S_{x}$ (b) $S_{x}$ (c) $S_{x}$ respectively.}
\end{figure*}

Unfortunately this extrinsic dependence of the Hall
conductivity on the applied field led to the introduction of an unphysical concept of an \textquotedblleft Anomalous Hall effect\textquotedblright%
\ (AHE), intrinsically proportional to the field-induced magnetization \cite{Pugh},
\begin{equation}
R_{H}=a^{O}H+a^{A}M,\label{OA}%
\end{equation}
where the coefficient $a^{A}$ was assumed to be a constant. Since this formula is unphysical and cannot be
applied to a single domain sample, the deviation from Eq. \ref{OA} have been routinely observed experimentally and dubbed by authors as either \textquotedblleft topological HE\textquotedblright%
\ (THE)\cite{PhysRevLett.93.096806}, \textquotedblleft antiferromagnetic HE\textquotedblright, or \textquotedblleft crystal HE\textquotedblright.\\

While the above are very good and insightful research dissemination, they focus primarily on the violation of linear dependence on $M$ (often to the extend that the HE appear at zero $H$ and zero $M),$ and provide a picture where
Eq. \ref{OA} is microscopically valid, although there are various esoteric exceptions. As a result experimentalists have worked out an intrinsically flawed procedure that is routinely applied to magnetotransport measurements, whereupon an attempt is made to fit the experimentally measured Hall resistivity to the
modified Eq. \ref{OA}, namely
\begin{equation}
R_{H}(H)=a^{O}H+a^{A}M(H)+R^{T}(H),\label{OAT}%
\end{equation}
where the coefficients $a$ are assumed to be field- and moment-independent.
The superscript $T$ here stands for \textquotedblleft
topological\textquotedblright,\cite{Gerber} since this is the most common designation of this residual term.\\

A microscopically consistent definition of AHE was provided in numerous publications where the AHE arises from the off-diagonal conductivity not
resulting from the Lorenz force, and expressed, via the Kubo formalism\cite{Bruno, Qian}, as%
\begin{align}
\sigma_{xy}^{A}  & =\sum_{n\neq m}\int d^{3}\mathbf{k}[f(\epsilon
_{kn})-f(\epsilon_{km})]\\
& \times\operatorname{Im}\frac{\left\langle \mathbf{k}n|\hat{v}_{x}%
|\mathbf{k}m\right\rangle \left\langle \mathbf{k}m|\hat{v}_{x}|\mathbf{k}%
n\right\rangle }{(\epsilon_{kn}-\epsilon_{km})}%
\end{align}
where the integration is properly normed, and the velocity operator is defined as $\mathbf{\hat{v}=}\partial\mathcal{H}(k)/\partial\mathbf{k,}$ the derivative of the Hamiltonian. Here we omit dissipative terms and $f(\epsilon)$
is the Fermi distribution function. Nowadays a more elegant formulation in terms of the geometrical Berry curvature\cite{Xiao} is more often used, namely
\begin{align}
\sigma_{xy}^{A}  & =-\sum_{n}\int d^{3}\mathbf{k}f(\epsilon_{kn})B_{n}%
^{z}(\mathbf{k)}\\
\mathbf{B}_{n}(\mathbf{k)}  & \mathbf{=}i\frac{\partial}{\partial\mathbf{k}%
}\times\left\langle \mathbf{k}n|\frac{\partial}{\partial\mathbf{k}}%
|\mathbf{k}n\right\rangle .
\end{align}
\\
There are various mechanisms that render nonzero values for this expression, including, but not limited to, uncompensated collinear magnetism (\textquotedblleft classical\textquotedblright\ AHE) \cite{Hurd}, noncollinear co-planar antiferromagnetism\cite{PhysRevLett.112.017205}, compensated
collinear magnetism (so-called altermagnetism)\cite{smejkal2021altermagnetism}%
, and non-coplanar spin-chiral magnets\cite{Machida2010}. In this didactical paper we will show
that deviations from this Equation may arise in absence of any esoterical mechanism, through trivial dependence of electronic structure on magnetic order.\\
\begin{figure*}[pt]
	\centering
	\includegraphics[width=0.8\linewidth]{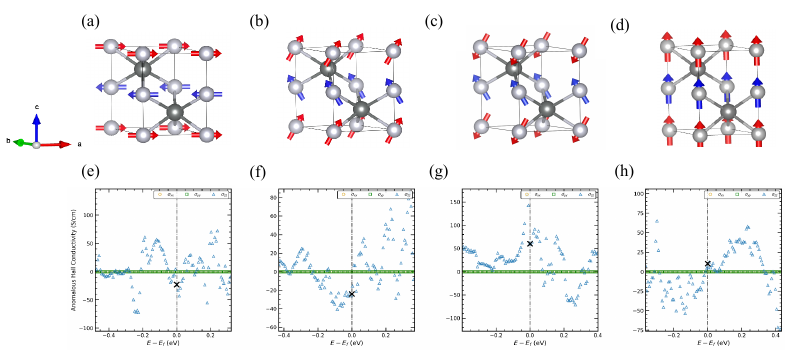}
	\caption {The four magnetic configurations with (a) coplanar (b) canted at 45$^{\circ}$ (c) canted at an angle of 135$^{\circ}$ and (d) collinear orientations of the spin moment on the Cr sublattices (e)-(h) Anomalous hall conductivity  as a function of the shift in the fermi energy for the four configurations.}
\end{figure*}

\begin{figure}[pt]
	\centering
	\includegraphics[width=1.0\linewidth]{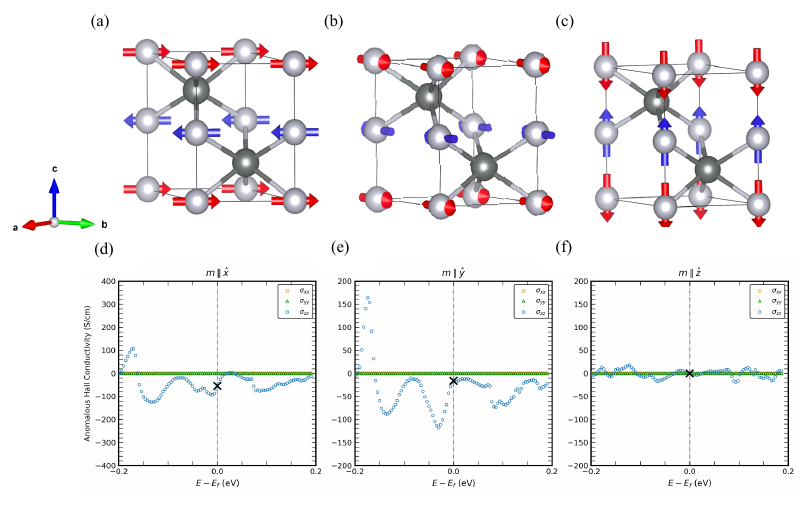}
	\caption {The three magnetic configurations with spin moments on the Cr aotoms opposite to each other along (a) a axis (b) b axis (c) c axis respectively resulting in zero magnetic net moment. (d)-(f) Anomalous hall conductivity  as a function of the shift in the fermi energy for the three configurations.}
\end{figure}

\begin{figure}[pt]
	\centering
	\includegraphics[width=0.7\linewidth]{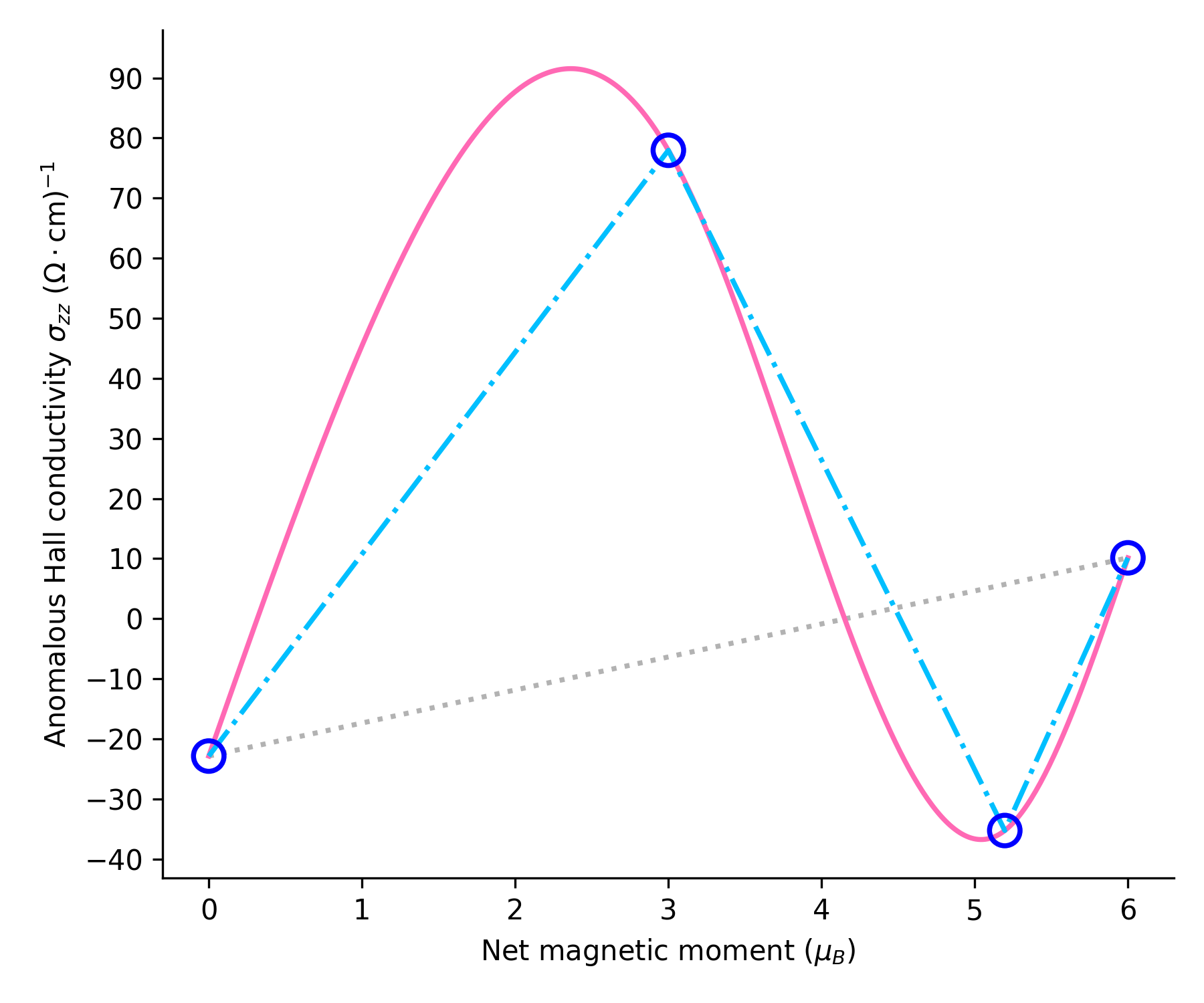}
	\caption {Plot of the anomalous hall conductivity as a function of the net magnetization for four different magnetic configurations for the altermagnetic compound CrSb.}
\end{figure}

\section{Results and Discussions}

In this section, we display the spin and orbital resolved electronic structure, phonon dispersion, spin orientation and topological properties such as surface states and geometrical berry curvature of the altermagnetic compound CrSb from first principles calculations\cite{Giannozzi_2017, kresse1996} combined with maximally localized wannier functions\cite{Vanderbilt}. We have investigated the electronic structure, phonon dispersion, magnetic properties and topological features using first principles Density functional theory (DFT). DFT calculations were carried out using  the projector-augmented wave method implemented in the Vienna ab initio Simulation Package (VASP)\cite{kresse1996} and Quantum espresso\cite{Giannozzi_2017} codes. The exchange-correlation functional form of the generalized gradient approximation implemented within the PBE functional was leveraged for our first principles calculations. In our ground state calculations, plane wave basis sets were utilized with energy cutoff of 600 eV and convergence was achieved with a 8$\times$8$\times$8 k-point grid.
The relaxation of the internal coordinates were performed while keeping the cell shape fixed using the VASP code until the ellmann–Feynman forces reached the threshold of 0.001 eV/$\AA$. The ion steps of the converged with a threshold of 10$^{-6}$ eVwhereas the self-consistent electronic steps were converged upto a threshold of 10$^{-8}$.\\

Fig. 1 illustrates the zero magnetization due to the alternating spin sub lattices as dictated by the crystal symmetry $[E\parallel H] \: + \: [C_{2}\parallel AH]$ depicted along the $L_{1}-\Gamma-L_{2}$ direction. CrSb crystallizes in the P6$_{3}$/mmc-D$_{6h}^{4}$ space group\cite{Libor2022} with hexagonal symmetry. The non-trivial spin sublattice $[E\parallel H]$ corresponds to symmetry transformations responsible for the exchange of atoms belonging to one of two spin sublattices, while the other half of transformations in real space are dictated by the spin group symmetry $[C_{2}\parallel AH]$\cite{Libor2024}. The nature of anisotropies of the sublattice spin densities and the individual spin resolved fermi surfaces are characterized by these symmetries. The spin splitting from the two sublattices alternates between up to down spin across the entire Brillouin zone (BZ) leading to a net zero magnetization.\\

Fig. 1(a) and (b) displays the spin resolved (up and down respectively) bands along the $L_{1}-\Gamma-L_{2}$ direction, whereas Fig. 1(c)-(f) displays the orbital and spin resolution of the electronic bands. Note that the Cr $3d$ orbitals contribute significantly towards both the valence and conduction states of the electronic spectrum, whereas the Sb $5p$ orbitals contribute significantly to the conduction states respectively. A detailed comparison reveal that our DFT calculations utilizing VASP and Quantum espresso captures the features of the electronic spectrum and spectral properties in reasonably good agreement with that of experimentally measured Angle-resolved Photoemission Spectroscopy (ARPES) findings\cite{Reimers2024, Yang2025}. Figure 2(a) exhibits the phonon spectrum of the altermagnetic compound along with the density of states resolved by the atomic decomposition (Fig. 2 (b). Note that the phonon spectrum is painted by the inverse participation ratio (IPR)\cite{Jarvist} of the phonon modes. The calculated IPR contributions hint towards the real-space localization of the phonon modes in the altermagnetic compound. We observe that the lowest energy (highly dispersive) acoustic phonon modes present the lowest IPR rates, similar to that of the (highly localized) high energy optical modes. Fig. 2(c) exhibits the fractional energy participation of the sublattice atoms (Sb and Cr atoms) revealing a significantly higher energy participation from the Cr (Sb) atoms for the optical (acoustic) modes compared to the acoustic (optical) modes respectively. This is further confirmed by the phonon density of states showing alternate participation of the Cr and Sb atoms at the high frequency limit (5-8 THz) and lower frequency range (0-4 THz) respectively of the phonon spectrum. \\

Next, we have utilized DFT to investigate the momentum and band resolved geometrical berry curvature contributions and possibility of surface states in CrSb. The formalism involves the generation of maximally localized wannier functions from first principles density functional calculations. In order to perform these calculations, we have carried out a unitary transformation from the Bloch states to obtain localized Wannier functions. Here, the degree of freedom provided by the indeterminacy of the phase of the Bloch wavefunctions have been utilized for the construction of \textquotedblleft{maximally localized}\textquotedblright Wannier functions. An accurate convergence of the minimum spread of the wannier functions is crucial for the above step. Following our wannier interpolation scheme, a mixed Fourier transform of the regular grid of $\textbf{k}$ points in addition with recursive adaptive mesh iterations have been performed utilizing the WannierBerri code further in order to obtain berry curvatures with fine k-meshe grids of 200$\times$200$\times$200 without significant computational resources. The accuarte berry curvature contributions are obtained by a combination of the VASP interface of Wannier90\cite{Vanderbilt} and WannierBerri codes\cite{Tsirkin2021}.\\

Fig. 3 (a),(b) exhibit the fermi surface of CrSb colour-coded by the geometrical berry curvature contribution across the entire BZ (red and blue for positive and negative berry curvature contributions respectively).  Fig. 3(c) displays a quiver plot indicating the distribution of berry curvature in the material along the ab plane, providing computational estimates of the topological phases, anomalous hall coefficient and nontrivial band crossings. Note that the berry curvature contributions from the quiver plots in the ab plane mimics the berry curvature resolved fermi surfaces when onserved from the c axis depicted in Fig. 3(b). Fig. 3 (d)-(f) present contribution to the geometrical curvature from each of the individual electronic bands crossing the fermi surface and reflects the  symmetry properties characteristic of altermagnetic compounds.\\

During the course of this study, we came across recent experimental ARPES publications that identified topological surface states in the electronic properties of this materials\cite{Lu}. Motivated by these experimental findings, we have utilized the maximally localized wannier functions (MLWF)\cite{Vanderbilt} for the construction of a semi-emperical tight binding (TB) model for evaluating the topological calculations in CrSb. In an event, the topological bulk bands are non-trivial, topologically protected surface states are obtained. The WannierTools\cite{WU2018405} code was utilized for the construction of tight binding model from the wannier functions for the estimation of the existence of surface states in the material. We have constructed a slab system for our calculations of surface states with 30 slabs (periodic along the two surfaces of the slab)\cite{WU2018405}. The solution of the Hamiltonian of the slab system provides the solutions of the surface states in the material. Figure 4 and 5 provides the spectral features of the surface states from our first-principles and tight binding calculations. Fig 4(a) shows the computed surface density of states and electronic structure for bulk CrSb near the fermi level, whereas Fig. 4(b)-(c) provide the surface density of states and surface state (SS) spectrum respectively for the slab along the high symmetry $\boldsymbol{k}$ path\cite{WU2018405, Vanderbilt}. Note that, following our calculations fermi arcs were obtained from the ends of the six connected fermi surface segments at the periphery (similar to the six spokes of a wheel). Figure 5 depicts the surface spin resolved density of states (DOS) for the slab system, where Fig. 5 (a), (b) and (c) displays the Pauli matrices $\boldsymbol{S_{x}}$, $\boldsymbol{S_{y}}$ and $\boldsymbol{S_{z}}$ respectively evaluated by computing the spectral functions of the surface Green's functions. Note that the DOS as a function of energy (Figure 4)  and spin resolved DOS (Figure 5) are shown as observed from the top of the slab geometry (c axis). The signatures of symmetry characteristic of altermagnets are observed from the plots of the spin resolved density of states in conformity with the electronic structure of Figure 1 and previously published results.\cite{Libor2022, Libor2024}\\

In the final part of the dissemination, we address the controversy regarding the linear behavior of AHC and the justification for assigning the disgression from linearity to topological Hall effect. In order to validate the linearity of the dependence of AHC\cite{Ong} with the net magnetization in the case of an altermagnet, we have simulated the effect of applying magnetic fields in the altermagnetic material CrSb\cite{smejkal2021altermagnetism}. In order to do so, the magnetization on the two different Cr sublattices, have been considered separate from each other (except one scenario) for different directions of magnetic moments in response to applied external fields. In Figure 6, we depict four orientations of the sublattice spin moments and the corresponding AHC for the test material CrSb. Fig. (a)-(d) shows configurations where the spin orientations are coplanar (a), collinear and parallel (d), staggered at $45^{\circ}$ (b) and staggered at $135^{\circ}$ respectively. Note that the first principles noncollinear calculations are performed by constraining the spin moments along these directions in the Cr sub-lattices, followed by construction of maximally localized wannier functions, mixed fourier transforms and adaptive mesh iterations for computation of the berry curvature contributions for each of the configurations above. The berry curvature contributions from all the electronic bands over the entirety of BZ integrated upto the fermi level provides an accurate estimation of the intrinsic AHC contribution in the material\cite{Tsirkin2021}. Fig. 6 (e)-(h) shows the AHC contributions for the four different configurations of CrSb with small variations around the fermi level $E_{f}$. An intersection of the plot with $E\:=\:E_{f}$ provides an estimate of the value of AHC as illustrated by the black cross in the figures.\\

A similar exercise has been shown in Figure 7, where the spin moment on the Cr sublattice (blue) is oriented along the a, b and c directions respectively (Fig. (a)-(c), whereas the spin moment on the other sublattice is in the opposite direction i.e. $-$a, $-$b and $-$c directions leading to two coplanar and one collinear configuration. Note that in all three scenario, there is a zero net magnetic moment from the Cr atoms in the material. First principles DFT calculations aided by wannier interpolations and adaptive mesh iterations utilizing WannierBeeri code, result in different AHC contributions from all three cases (nonzero from two, zero from one).\\

For the purpose of visual illustration, the dependence of the AHC on the net magnetic moment\cite{Niu} from the four configurations in Fig. 5 have been displayed in Figure 7. We note that, for the altermagnetic compound CrSb hosting topological features in the electronic spectrum, the AHC doesn't depend linearly on the magnetization, suggestive of the fact that the previous theory of linear dependence of AHC is valid, at best, for experiments on a subset of magnetic materials with single magnetic domains. We find it to be unjustified to extrapolate the linear dependence of AHC on magnetization to include all classes of collinear and non-collinear magnets, anti ferromagnets and altermagnetic compounds\cite{Igor}.

\section{Summary and Conclusions}

In summary, we have investigated the classification of Hall effect into the didactic classification of OHE, EHE and AHE, and the misnomer of generalization of deviations from the linearity of AHE with topological Hall effect (THE). A detailed first principles DFT calculation and tight binding model constructed from Wannier functions of the electronic spectrum, and topological properties of CrSb has been presented. Our results indicate symmetry enforced momentum and band resolved geometrical berry curvatures and surface states originating from nontrivial topological signatures in the altermagnetic compound. The spin resolved density of states from construction of slabs along the (001) surface utilizing tight binding model has been discussed. Finally, quantum simulations of two different set of configurations, with zero and non-zero net magnetic moment have been investigated. First principles calculations of AHC in these configurations reveal the fact that the altermagnetic compound oriented in external fields fail to concur with the intrinsically flawed\cite{Niu, Mazin2026} hypothesis of linear behaviour of AHC with net magnetic moment, inspite of being routinely applied experimentally to all classes of magnetic materials. During the summarization of this study, we came across a similar insightful dissemination, where the recipe of linear dependence was found to be in disagreement for the weak itinerant ferromagnet\cite{Mazin2026}. In the future, our goal is the eprformance of detailed investigation of a broader class of altermagnetics, and envisage a prescription that works for altermagnetic compounds. 

\section{Acknowledgements}
The authors would like to thank Igor Mazin and Stepan Tsirkin for valuable discussions, suggestions, particularly for the methodology of the paper. The calculations were performed at the FESC computational facility, George Mason University and the Sharanga high performance computing facility, Birla Institute of Technology Hyderabad.


\bibliography{CrSb.bib}

\end{document}